\documentclass[10pt]{article}
\usepackage{amsmath}
\usepackage{amssymb}
\usepackage{graphicx}
\usepackage{hyperref}

\begin{document}

\setcounter{figure}{0}
\setcounter{table}{0}
\setcounter{footnote}{0}
\setcounter{equation}{0}

\vspace*{0.5cm}

\noindent {\Large MEASURES OF THE EARTH OBLIQUITY DURING THE 1701 WINTER SOLSTICE AT THE CLEMENTINE MERIDIAN LINE IN ROME}
\vspace*{0.7cm}

\noindent\hspace*{1.5cm} A.H. ANDREI$^{1,2,3,4}$, C. SIGISMONDI$^{1,3,5,6}$, V. REGOLI$^6$\\  
\noindent\hspace*{1.5cm} $^1$ Observat\'orio Nacional/MCTI\\
\noindent\hspace*{1.5cm} Rua Gal. Jos\'e Cristino 77, Rio de Janeiro, RJ CEP 20921-400, Brasil\\
\noindent\hspace*{1.5cm} e-mail: oat1@ov.ufrj.br\\
\noindent\hspace*{1.5cm} $^2$ SYRTE/Observatoire de Paris\\
\noindent\hspace*{1.5cm} Avenue de l'Observatoire 61, Paris 75014, France\\
\noindent\hspace*{1.5cm} $^3$ Observat\'orio do Valongo/UFRJ\\
\noindent\hspace*{1.5cm} Ladeira do Pedro Ant\^onio 43, Rio de Janeiro, RJ CEP 20080-090, Brasil\\
\noindent\hspace*{1.5cm} $^4$ Osservatorio Astrofisico di Torino/INAF\\
\noindent\hspace*{1.5cm} Strada Osservatorio 20, Pino Torinese, TO 10025, Italia\\
\noindent\hspace*{1.5cm} $^5$ Istituto Tecnico Industriale Galileo Ferraris\\
\noindent\hspace*{1.5cm} Via Fonteiana 111, Roma CAP 00152, Italia\\
\noindent\hspace*{1.5cm} e-mail: costantino.sigismondi@gmail.com\\
\noindent\hspace*{1.5cm} $^6$ Pontifical Athenaeum Regina Apostolorum\\
\noindent\hspace*{1.5cm} Via degli Aldobrandeschi 190, Roma CAP 00163, Italia\\

\vspace*{0.5cm}

\noindent {\bf \large ABSTRACT.} 
The great meridian line in the Basilica of Santa Maria degli Angeli in Rome was built in 1701/1702 with the scope of measuring the obliquity of the Earth's orbit in the following eight centuries, upon the will of Pope Clement XI. During the winter solstice of 1701 the first measurements of the obliquity were taken by Francesco Bianchini and here they are firstly published. He was the astronomer who designed the meridian line, upgrading the similar instrument realized by Giandomenico Cassini in San Petronio, Bononia. The accuracy of the data 
is discussed from the point of view of the original setup of the astrometric pinhole.
\vspace*{1cm}

\noindent {\bf \large 1. THE ASTROMETRIC PINHOLE}

\smallskip

All ancient meridian lines have been re-measured after some decades of duty, in order
to verify their alignment and the position of the pinhole. These instruments have
been built to measure the variation of the obliquity along the centuries, and the need
of a re-calibration was part of the observational duties. The Cassini meridian line in
San Petronio, Bologna, made in 1655 was revised in 1695 by the same astronomer
Giandomenico Cassini. Similarly Leonardo Ximenes in 1761 restaured the meridian
line in Santa Maria del Fiore in Florence, made by Paolo del Pozzo Toscanelli
in 1475. The great Clementine gnomon of Santa Maria degli Angeli in Rome,
completed by Francesco Bianchini in 1702, was studied and remeasured by Anders
Celsius in 1734 and Ruggero G. Boscovich in 1750. They found the deviation of
the azimuth from the true North, respectively of 2' (1734) and 4'30" (1750). Our
measurements of 2006 (Sigismondi, 2013), used the Polaris' transits technique, yielding 4' 28".8 $\pm$ 0.6",
in agreement with the measurements made by Boscovich.

In the recognitions of Cassini and Ximenes the main issue was the movement of
the pinhole with respect to the original position.

This was due to the fact that the pinhole in Bologna was on the roof, and in
Florence was in the dome of the church: both positions were subjected to motions
of the buildings due to thermal response, winds and settling of the walls.

For this reason Francesco Bianchini chosen the basilica of Santa Maria degli
Angeli in Rome to build the meridian line upon the will of Pope Clement XI (1700-
1721): this church was built by Michelangelo in the original roman hall of the Diocletian
baths, a 1500 years old structure, with no more settling ongoing.

\newpage

\noindent {\bf\large 2. RESULTS}

\smallskip

\begin{figure}[h]
\begin{center}
\includegraphics[scale=1]{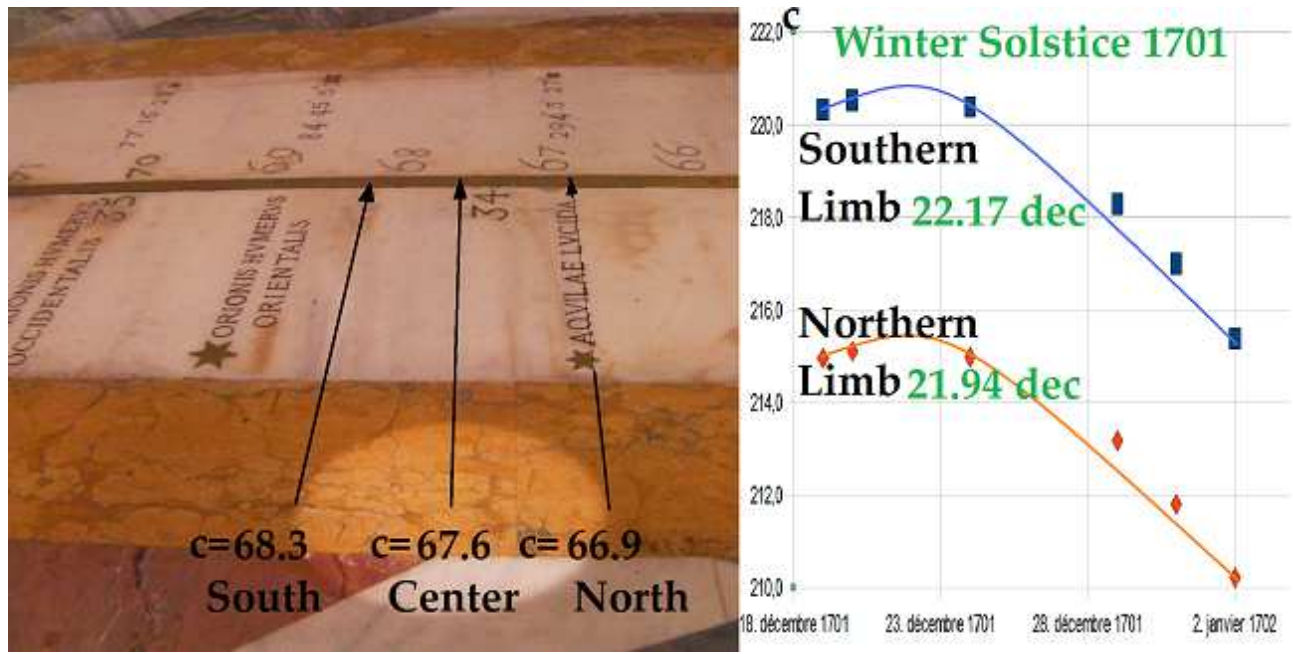}
\caption{The image of the Sun is projected through a pinhole on the floor moving up on September 2$^{nd}$ 2014. The hundredth part "c" of the pinhole's height is the unit of measurement along the meridian line, in such a way $c=100\times tan(z)$. The solstitial data reported in the letter of Francesco Bianchini to Pope Clement XI are plotted on the right hand side of the figure as the Southern and Northern limb positions.}
\end{center}
\end{figure}

On the left part of Figure 1 we can see that the center of the Sun has $\it{c}$ $\sim$ 67.6 in remarkable agreement with the \href{http://vo.imcce.fr/webservices/miriade/rts_query.php?-body=11&-nbd=1&-long=-12.4961167&-lat=41.9024223&-ep=2014-09-02&-from=Cassini&-mime=text&-extrap=4}{IMCCE/Observatoire de Paris ephemeris for the day} the zenit angle of the solar center $\it{z}$=34$^\circ$.0548, equivalent to $\it{c}$=67.590. Bianchini could measure the position of the center of the Sun to the nearest arcsecond by drawing both the locations of the Southern and the Northern limbs of the Sun.

On the right part of Figure 1 the quadratical fit with the data, never published up to now, of the Southern limb yields 22.17 December 1701 for the solstice, and 21.94 December fitting with the Northern limb data;  averaging N and S limbs we obtain the solstice on 22 December 1701 at 00:26 UT, while the \href{http://www.imcce.fr/en/grandpublic/temps/saisons.php}{IMCCE ephemeris} for 1701 give 21 December at 23:35 UT.

The same quadratic fit yields for the extreme positions of the two limbs of the Sun at the solstice time: Southern 220.597 and Northern 215.228.
Correspondingly the unperturbed center of the solstitial Sun has declination $\delta$=23$^\circ$28'48", being 41$^\circ$54'11".2 the latitude of the pinhole.

Thus the observed mean Obliquity can be derived as $\epsilon$=23$^\circ$28'54".3, after considering the \href{http://www.neoprogrammics.com/nutations/}{nutation phase} for that time.
This is in excellent agreement with modern calculations for the mean obliquity in 1702.0: Laskar method gives $\epsilon$=23$^\circ$28'40".9, whereas Duffett-Smith method gives $\epsilon$=23$^\circ$28'58".6.

These measures confirm the accuracy achieved in the original setup of the Clementine Meridian line. 
\smallskip
\smallskip

\noindent {\bf\large 3. REFERENCES}

{

\leftskip=5mm
\parindent=-5mm

\smallskip
\begin{itemize}
\item{Bianchini, F.; 1702; Letter "To The Holiness of Our Lord Pope Clement XI", Biblioteca Vallicelliana, Fondo Bianchini, Roma}
\item{Sigismondi, C.; 2013; "The astrometric recognition of the Solar Clementine Gnomon (1702)"; International Journal of Modern Physics, Conference Series, vol 23, 454.}
\item{Laskar, J.; 1986; "Secular terms of classical planetary theories using the results of general theory"; Astron. Astrophys. 157, 59-70.}
\item{Duffett-Smith, P.; 1990; "Astronomy with your personal computer", Cambridge University Press, Cambridge, UK.}
\end{itemize}
\end{document}